Proposal of Absolute Nanometer Size Measurement in Flow Cytometry
Based on Laser Interferometry

Tsumoru Shintake
OIST: Okinawa Institute of Science and Technology Graduate University
1919-1 Tancha, Onna, Okinawa 904-0495 Japan
e-mail: ShintakeLab@oist.jp

Abstract

The author discusses the laser interference method to measure the size of small bio-particles: extracellular vesicles (EVs), exosomes and viruses of nanometer scale in flow cytometry. By introducing a new laser configuration in place of conventional optical system of the flow cytometry, the interference fringe (periodic intensity modulation) is created inside the flow which provides a calibrated scale for the size measurement. The fringe pitch is precisely determined by the crossing angle of laser beams and the wavelength. The transverse size of the beam spot can be much larger than wavelength, for example, 30 micro-meters, where the flow cell design in conventional system can be used with moderated dimensional tolerance.

The interaction length of sample with laser becomes longer: ~30 micro-meter and thus the scattering light and fluorescent light in case of labeled particle becomes much higher. Importantly, those signals are modulated in MHz frequency range associated with the periodic intensity modulation of the interference fringe. By measuring modulation depth, we can determine the particle size. We may utilize a bandpass filter or the digital signal processing of FFT, and thus we can drastically improve S/N ratio. This is because of the narrow spectrum of signal and thus effectively eliminates noise from various debris contained in the sample, and also low frequency shot noise in the photo detector signal. Thanks to the above-mentioned merits, it will be possible to measure the size of EVs and exosomes in particle-by-particle basis in reliable dimensional scale at nano-meter resolution. It will be also possible to realize the fast virus detection system for screening passengers in the airport at the current pandemic of COVID-19 and in future.

1. Introduction

Flow cytometry is a general method for rapidly analyzing large numbers of cells individually using light-scattering, fluorescence, and absorbance measurements. The recent success of flow cytometry is based on commercially available flow cytometry equipment that is both robust and versatile, together with modern data acquisition and analyzing software, and progressed various staining fluorescence assays [1].

However, the size measurements are highly influenced by the refractive index, and also practically limited to 300 nm size or larger. Below this size the scattered light intensity drops quickly and thus it becomes rather difficult to determine the size, while there are physiologically important particles to be studied: extracellular vesicles and viruses.

All body fluids contain cell-derived membrane-enclosed vesicle; extracellular vesicles (EVs). Such vesicles are emitted by bio-cells and contain messages to the other organs [2]. The size characterization of extracellular vesicles (EVs) and drug delivery liposomes is of great importance in their applications in diagnosis and therapy of diseases [3]. Figure 1 shows conceptual diagram of size distribution on EVs and viruses. The vertical axis is the expected Rayleigh scattering photons in a laser beam, which is calibrated arbitrarily; 100 photons at 10 nm size.

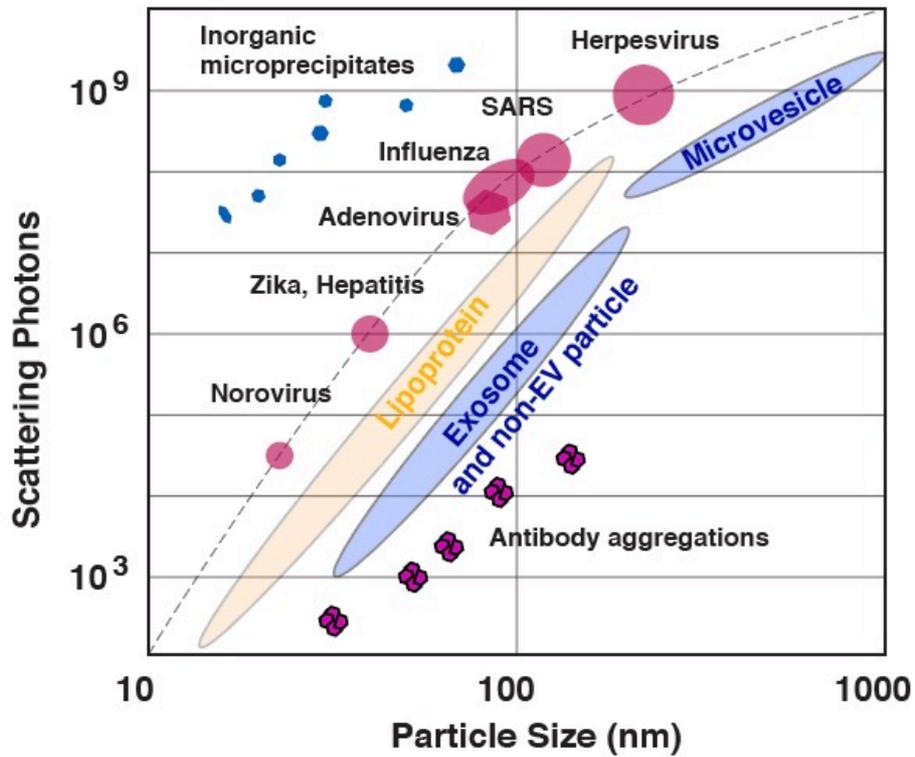

Fig. 1 Conceptual diagram of size distribution on EVs and viruses.

The Rayleigh scattering cross-section is given by

$$\sigma_s = \frac{2\pi^5}{3}\frac{d^6}{\lambda^4}\left(\frac{n^2-1}{n^2+2}\right)^2 \approx 90\,\frac{d^6}{\lambda^4}\delta n^2 \qquad (1)$$

Where $d$ is the particle size, and $\delta n$ is refractive index difference its medium (water).

The cross-section scales as 6th-power of size, i.e., smaller particle does not scatter enough photons. In Fig. 1, the scattering photon quantity changes $10^6$ times from 10 nm to 100 nm. Above 200 nm, $d>\lambda/\pi$, the scattering behavior follows Mie-scattering theory, and finally approaches to the physical disk cross-section. In order to challenge small cross-section for small size, we may introduce the following improvements.

(1) Shorten the laser wavelength
(2) Increase laser power density
(3) Improve photon capture efficiency
(4) Increase the interaction length (longer interaction time)
(5) Reduce noise in the detection system
(6) Reduce noise source in the sample flow

There have been extensive R&Ds to extend the capability of flow cytometry toward nano-meter size [4, 5]. They mainly focused item (1), (2) and (3), by introducing high resolution objective lens having higher numerical aperture, while finer size of sample flow and tight alignment was required for the lens system. Some of them succeeded in observing nano-meter size plastic beads and could separate them according to their size. However, we have to note that the plastic beads have higher refractive index (polystyrene 1.65, silica 1.45) than actual EVs (1.36-1.40), and thus it is relatively easy to perform experiments [6].

Among above list, item (4) contradicts the aim of smaller size detection in the conventional system, i.e., small laser spot is desirable. In the proposed method in this paper, items (4) and (5) will be drastically improved by introducing the laser interferometry. Additionally, this new method provides absolute scale on the size measurement, and it does not require calibration by using artificial beads.

## 2. Introducing laser interferometry in the flow cytometry

Fig. 2 shows the schematic diagram of proposed method (right side), compared with conventional flow cytometry of split laser system (left side). In the conventional system[7], the laser beam splits in two parallel beams and illuminates on the flowing sample at two spots. The distance between them is known, and thus by the time-of-flight measurement between the two spots, the flow velocity can be calibrated. From the pulse width of sample crossing the upstream spot, we determine the sample size. The scattering signal has two pulses, accordingly, separating time $T$, and width $\delta T$. The particle size can be determined as follows.

$$d \approx \frac{\delta T}{T} L \qquad (2)$$

If the particle size is small, we have to correct effect of laser spot size. The detectable particle size is limited by laser spot size.

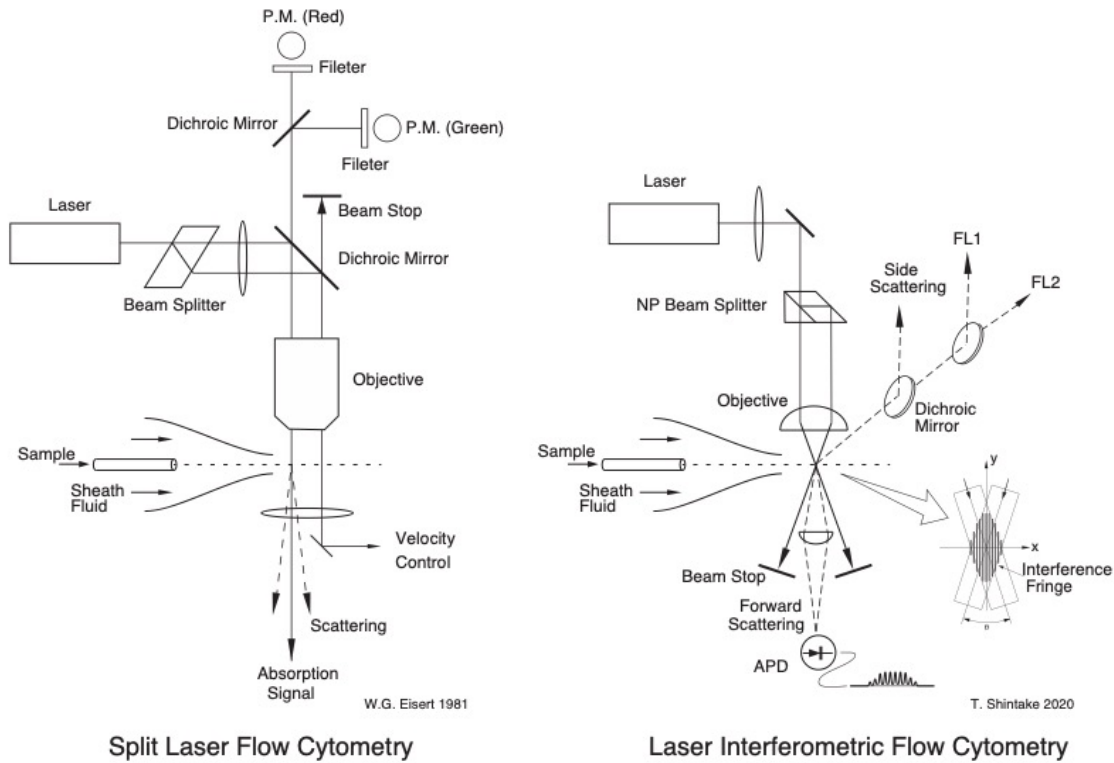

Fig. 2. (left) Conventional flow cytometry with split laser scheme. (right) This invention of laser interferometric flow cytometry.

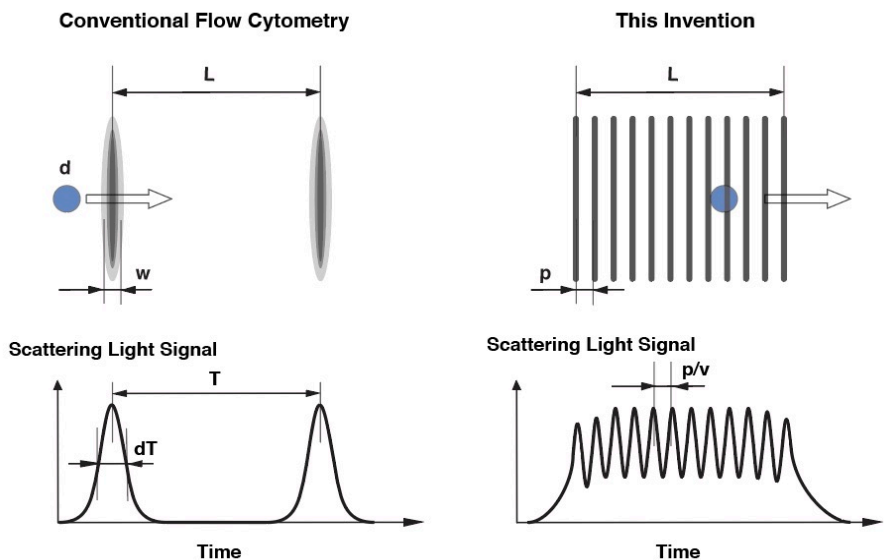

Fig. 3. Comparison of interaction region. (Left) Conventional flow cytometry, (Right) this invention using laser interference fringes for particle size determination.

The proposed method uses laser interference effect to determine size of the sample.

The laser beam splits into two parallel beams in a beam splitter, then focused into the flow cell.

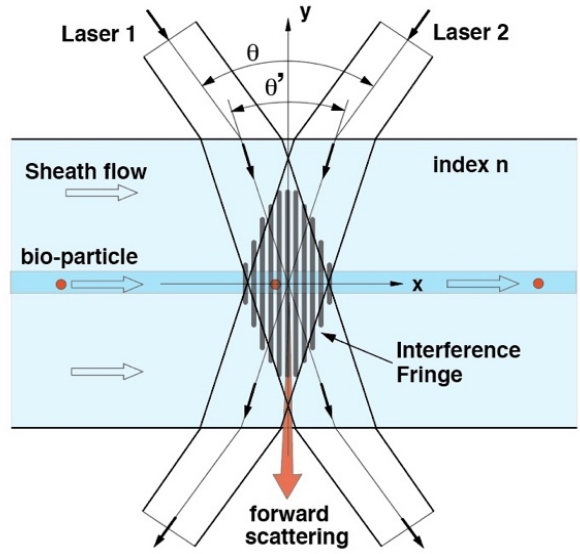

Fig. 4. Two laser beams crossing over the sample flow, create interference fringe.

As shown in Fig. 4, they overlap with opening angle θ', which create interference fringe of pitch $p$.

$$p = \frac{\lambda'}{2sin(\theta'/2)} \quad (3)$$

The wavelength in the flow is shorter than in air;

$$\lambda' = \frac{\lambda_0}{n} \quad (4)$$

where $n$ is the refractive index of the flow medium. The laser crossing angle becomes narrower in the medium, which is given by the Shell's law as follows.

$$\frac{sin\theta'/2}{sin\theta/2} = \frac{1}{n} \quad (5)$$

We have to note the critical angle for air-water interface is 48.6 degree (index of water 1.33), and thus we may not realize the laser crossing angle at this critical angle or higher.
Using eqs. (3), (4) and (5), we have

$$p = \frac{\lambda'}{2sin(\theta'/2)} = \frac{\lambda}{2sin(\theta/2)} \quad (6)$$

## 3. Photo-signals are frequency modulated and provide high S/N
The forward scattering power is periodically modulated at the carrier frequency;

$$f_0 = \frac{v_o}{p} \qquad (7)$$

where $v_0$ is the flow velocity, for example $v_0 = 1$ m/sec, we have $f_0 = 1.6$ MHz for $p = 638$ nm. This frequency is well within the response frequency of the high sensitivity APD (avalanche photo diode). The statistical noise in photo detector widely spreads in the frequency domain. Specifically, Poisson noise dominates at lower frequencies (1/f), so that we can effectively eliminate these noise signals by using narrow band-pass filter around the carrier frequency, and thus high S/N will be achieved.

There is a big advantage when we use staining fluorescence assays on the sample. The fluorescence light is also intensity modulated at the same frequency and therefore we may use a narrow band frequency filter to improve S/N ratio. We may simplify the wash process because fluorescent light from the solution simply creates low frequency bias signal which can be eliminated by the frequency filter.

## 4. Details of EM field of interference fringe and forward scattering light

The electro-magnetic (EM) field of the interference fringe is shown in Fig. 5.

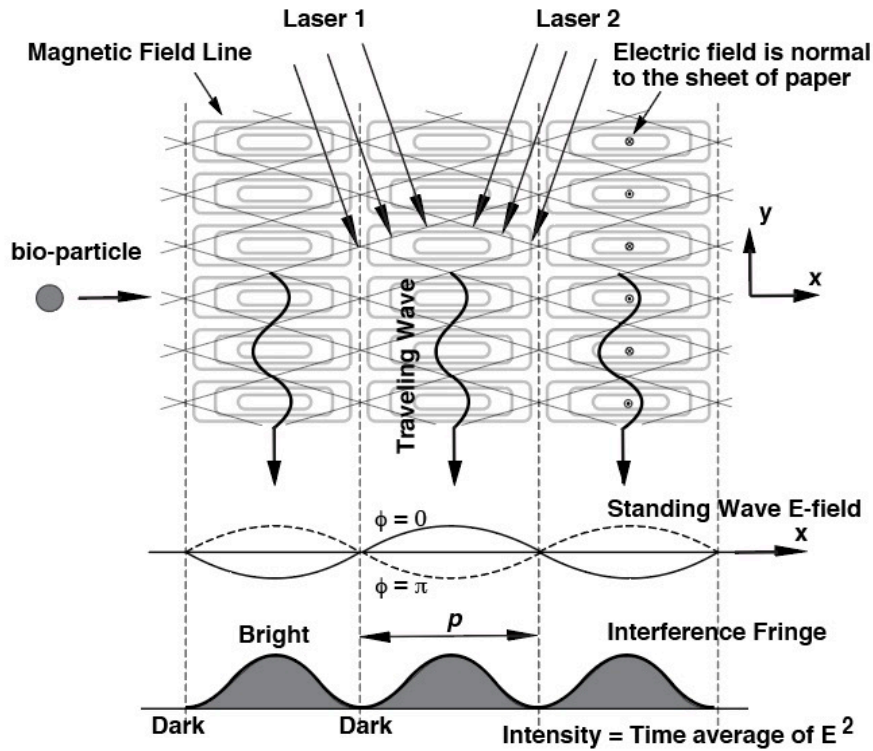

Fig. 5. EM field inside the interference fringe.

Two beam interference generates a standing wave pattern in x-direction, and traveling wave in y-direction, which is exactly same EM field inside the waveguide for the microwave circuit. The field intensity along x-direction has a standing-wave pattern, i.e., the time average takes peaks (bright zone) and zero (dark zone) repeatedly; this is the interference pattern. When a bio-particle is running through this pattern in x-direction, it feels the EM-field. Since the speed of the particle is much slower than the speed of the light, the contributions from the magnetic field becomes zero. The scattering light from a matter is, in fact, caused by scattering EM-field by the cloud of electrons (nuclei are too heavy to respond to optical frequencies and thus do nothing). The Lorentz force on electron is given by

$$\mathbf{F} = q(\mathbf{E} + \mathbf{v} \times \mathbf{B}) \approx q\mathbf{E} \qquad (8)$$

The electric field $\mathbf{E}$ is oscillating at optical frequency, and the electrons respond to this field; as a result, the electrons emit dipole radiation around them.

The light scattering from macroscopic object smaller than wavelength can be estimated by Rayleigh scattering. The scattering power has peaks at forward and backward directions, i.e., y-direction in Fig. 5. Total Rayleigh scattering cross-section is given in Eq. (1)

The scattering power from a particle inside the interference fringe is given by

$$P_s = \iint_S \sigma_s \, p_0(x, z) \, dx \, dz$$
$$= \iint_S \sigma_s \frac{E_0^2}{\zeta} \cos^2(k_x x) \, dx \, dz$$
$$= \sigma_s \frac{P_L}{\sigma_L} \int_{z_1}^{z_2} dz \int_{x_0 - d/2}^{x_0 + d/2} \frac{1 + \cos(2 k_x x)}{2} \, dx \qquad (8)$$

Where $P_L$ is laser power, $\sigma_L$ is laser spot size (cross-section), $x_0$ is the position of particle, $k_x$ is wave number in x-direction: $k_x p = \pi$, $z_1$ and $z_2$ is the integral limits representing circular particle. Equation 8 represents intensity modulation in the scattering light, which takes maximum at $2 k_x x_0 = 0$, and minimum at $2 k_x x_0 = \pm \pi$. Equation (8) is the Fourier transform of the particle shape and we can find the Fourier component at the modulation frequency by measuring scattering light from the particle in the interference fringe.

Here we define modulation depth as follows.

$$M = \frac{P_{max} - P_{min}}{P_{max} + P_{min}} \qquad (9)$$

By integrating Eq. (8), we find the modulation depth as follows (see Appendix 1)

$$M = \frac{k_2^2}{k_2^2 - (2 k_x)^2} \cos(k_x d)$$

$$= \frac{1}{1-(2d/p)^2} cos(\pi d/p) \qquad (10)$$

The modulation curve is shown in Fig. 6. By measuring amplitude modulation in the scattering light, we can deduce the particle size from this figure. If we assume our detection system can measure the modulation depth with 0.03 precision, the detection range becomes $M = 0.03 \sim 0.97$, the particle size can be seen at $d/p = 0.17 \sim 1.4$. For $p = 638$ nm, it becomes 108 nm ~ 890 nm. We have to note that two zones of $d/p = 1.3\sim1.5$ and $d/p = 1.5\sim2.5$ will produce same modulation depth. According to Eq. 1, two zones will have different light scattering power, thus we may distinguish them, easily.

The fluorescent light has the same modulation curve, it may slightly vary with different distribution inside the cell.

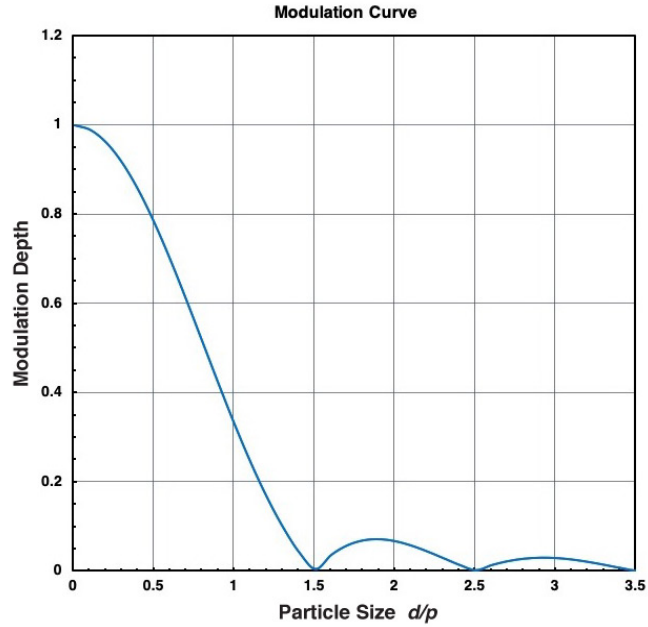

Fig. 6. The modulation depth as a function of the particle size.

As shown in Fig. 7, if the particle size is small, the light signal will be modulated very well, while if the size becomes larger and the light signal becomes larger, but the modulation becomes low. From the modulation measurement, we can determine the particle size by Eq. (10). Practical working ranges are summarized in Table-1 and 2.

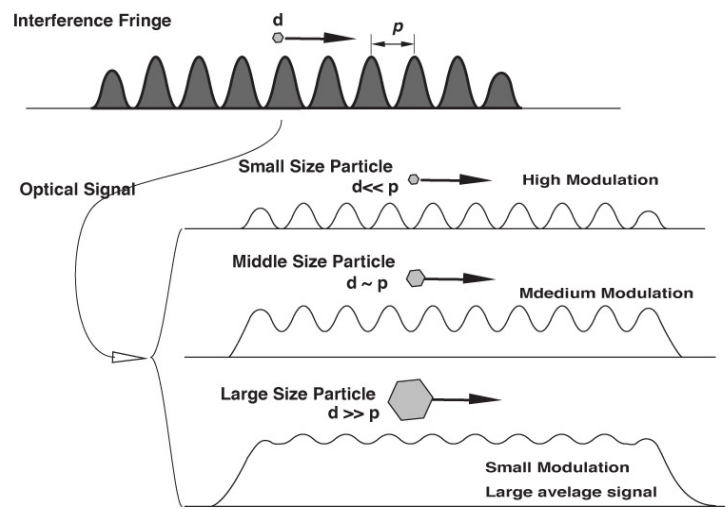

Fig. 7 Conceptual illustration of running particle in the interference fringe and generating light intensity.

Refractive index of water = 1.33

| Laser Type | Wavelength (nm) | Crossing Angle (deg) | Fringe Pitch (nm) | Working Particle Size (nm) |
|---|---|---|---|---|
| Green | 532 nm | 180 deg (cavity type) | 200 nm | 40 ~ 280 nm |
| -- | -- | 160 deg (prism coupling) | 203 nm | 41 ~ 284 nm |
| -- | -- | 120 deg (lens NA=1.2 water) | 230 nm | 46 ~ 322 nm |
| -- | -- | 30 deg (lens NA = 0.3) | 773 nm | 154 ~ 1082 nm |

Table-1 Working range using green laser (532nm)

Refractive index of water = 1.33

| Laser Type | Wavelength (nm) | Crossing Angle (deg) | Fringe Pitch (nm) | Working Particle Size (nm) |
|---|---|---|---|---|
| Ultraviolet | 355 nm | 180 deg (cavity type) | 134 nm | 27 ~ 189 nm |
| -- | -- | 160 deg (prism coupling) | 136 nm | 27 ~ 127 nm |
| -- | -- | 120 deg (lens NA=1.2 water) | 154 nm | 31 ~ 216 nm |
| -- | -- | 30 deg (lens NA = 0.3 ) | 518 nm | 103 ~ 725 nm |

Table-2 Working range using ultraviolet laser (355 nm)

Table-1, 2 show the working range for green and ultraviolet lasers. Refractive index of water 1.33 is assumed, and sensitive modulation depth of 0.05~0.95 is assumed.

A similar method was previously devised by author [9] and successfully demonstrated to measure the size of high energy electron beam [10]. In the course of R&D for future e+e- linear collider, FFTB test accelerator facility was constructed at SLAC Stanford Accelerator Center in 1990', and we focused 50 GeV electron beam down to 60~80 nm spot and measure its size by the interference fringe using 1.06 micron-meter wavelength YAG laser. For more detail refer to Appendix-2.

5. Practical Configurations

Figure 8 shows one of practical configuration, which uses objective lens and movable selection slit. The maximum crossing angle is limited by the numerical aperture of the lens and working distance, while this system makes it possible to operate as conventional cytometer with single spot, and also as the interferometric cytometer, by simply choosing slit shape. This system will be suitable to integrate into conventional cytometry.

Figure 9 shows prism coupling type, which provides higher crossing angle, while it requires additional work for two beam alignment to common focal point.

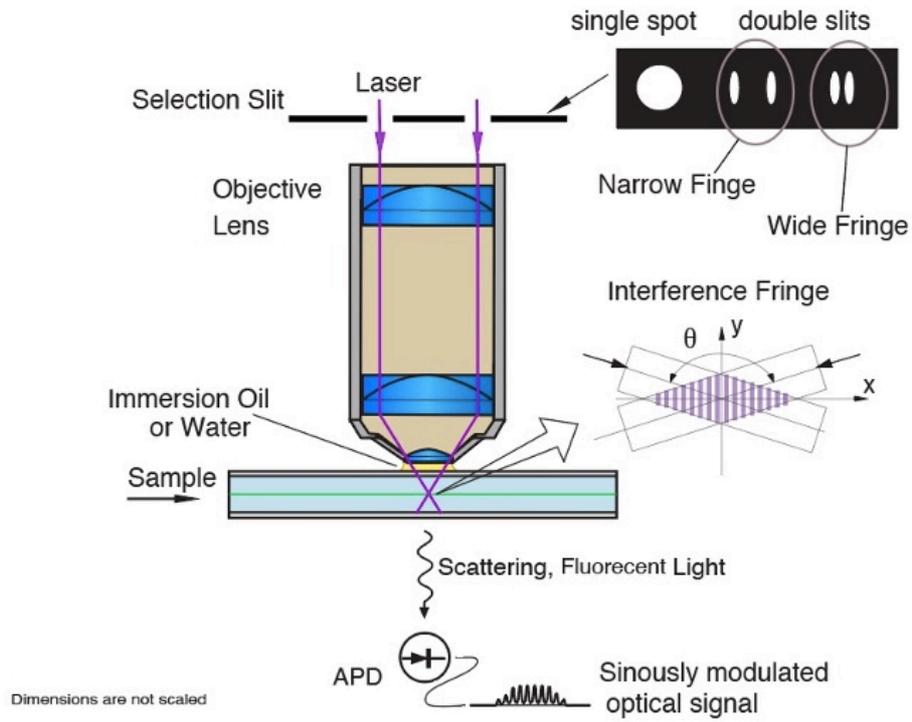

Fig. 8. Objective lens coupling configuration.

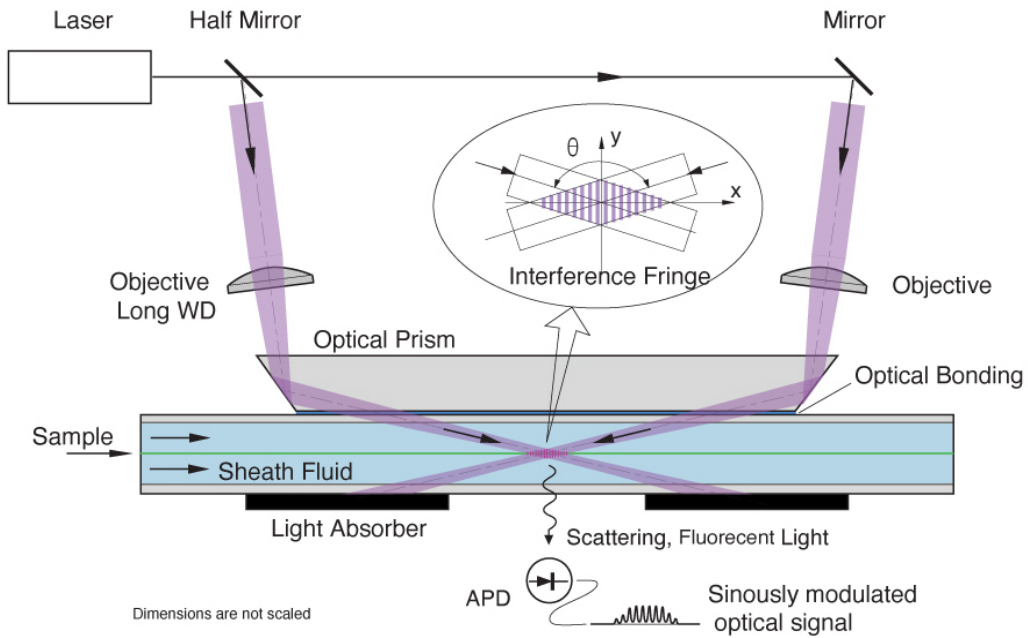

Fig. 9. Prism coupling configuration.

## 6. Cavity Resonator Type Flow Cytometry

Figure 10 shows cavity resonator type flow cytometry, which provides shortest interference pitch. The laser light enters into the cavity resonator through the partial reflecting cavity mirror-1, then is focused into the flow cell by the objective lens. By the 45-degree mirror, laser beam direction aligns to the flow cell, and focused to a spot, followed by reflecting by the downstream cavity mirror-2. Flow cell penetrates through the 45-degree mirror and the cavity mirror-2, thus some fraction of the laser power leaks from those holes, which will be fed by the laser source. The reflected power will be monitored through the isolator, and will be utilized to maintain the cavity resonance by feeding the error signal into piezo mover for the cavity mirro-1 position, and thus the laser intensity at the focus point will be stabilized. The circulating power inside the cavity may be much higher than the laser source power, which will provide more scattering photons.

Inside the cavity, there are two streams of waves, forward and backward waves, they overlap and create a standing wave, i.e., the interference fringes. The crossing angle becomes $\theta = \pi$, the fringe pitch becomes shortest,

$$p = \lambda'/2 = \lambda/2n \quad (11)$$

Using 488 nm blue laser, and n=1.33 for water, we have $p$ = 183 nm. We may detect particles within 30 nm to 260 nm size. According to Fig. 1, we may observe various viruses, including SARS coronavirus, and EVs.

Considering the field intensity at the waist of the laser focus, assuming the Gaussian beam, the Rayleigh length is given by

$$z_R = \frac{\pi w_0^2}{\lambda} \quad (12)$$

Where $2Z_R$ is waist length, w0 is the waist size. We should choose waist length long enough to obtain a large number of photons, while should be not too long to avoid multi-particle emission overlap. For example, if we chose $2Z_R$ = 30 micron-m, the waist size should be 1.5 micron-meter. The total angular spread of Gaussian beam becomes

$$\Theta_{div} \simeq 2\frac{w_0}{z_R} \quad (13)$$

which is 0.2 radian. If we chose the position of the 45-degree mirror and the cavity mirror-2 at 2 cm from the focus point, the Gaussian beam size becomes 4 mm in diameter, thus we should use mirror diameter 4 mm or larger. If we make a 1 mm hole for flow cell penetration, the leak power will be 2 x (1mm/4mm)$^2$ = 0.12, and thus the Q-factor of the cavity becomes 8, and the circulating power will be 8 times higher than the source laser power. The cavity finesse is ~47, and cavity tuning control will not be tight.

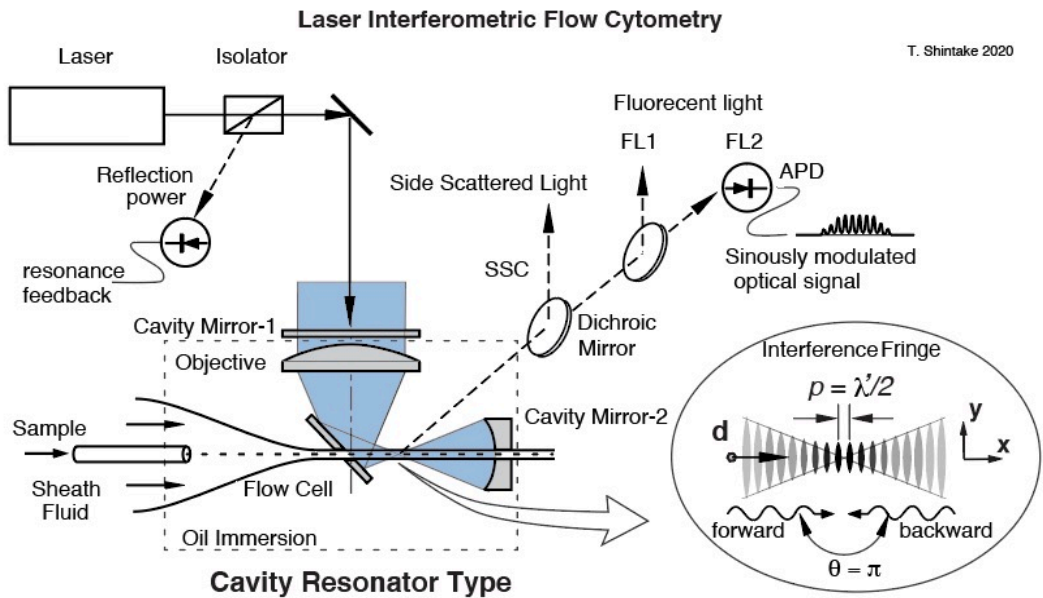

Fig. 10. Cavity resonator type flow cytometry.

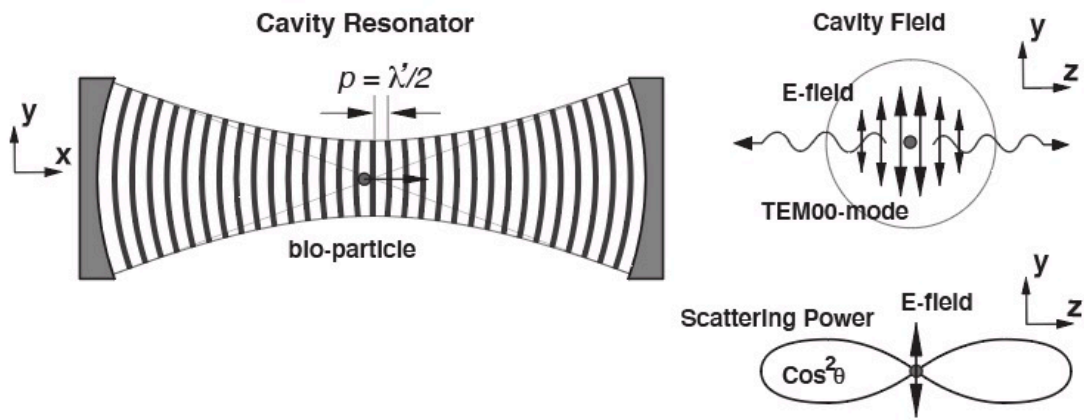

Fig. 11. EM field inside the cavity resonator.

Figure 11 shows the electro-magnetic field inside the cavity resonator. We assume dominant TEM00 Gaussian mode, thus field localized on axis. We may choose laser polarization as E-field in y-axis, and thus the magnetic field will be in z-axis. Because the speed of the particle is much slower than the speed of the light, thus the magnetic field does not do anything on the particle. The group of electrons on the particle is accelerated periodically at the optical frequency in y-axis, and emits dipole radiation, i.e., the Rayleigh scattering. The radiation pattern has the main lobe of $\cos^2\theta$ dependence, which is normal to the y-axis, and provides the side scattering (SSC) signal to the flow cytometry. In the cavity resonator type, the forward scattering (FSC) signal is not available.

## 7. Application to Fast Virus Detection System

In order to fight against the COVID-19 pandemic, we need a fast virus detection system. If we shorten the detection time to less than 10 seconds, for example, we may use this system to screen a large number of passengers in the airport and other key locations.

Because the proposed flow cytometry will provide a reliable scale on the particle size, which is robust against contaminations and another parameter change. Additionally, we should think to the nature; the structure of virus is unique and stable, round sphere of 100 nm in diameter, and highly packed with mRNA inside, resulting in a higher refractive index. The envelope mainly consists of the lipid bilayer, taken from the host cells at assembly/budding stage of the viral life cycle, and therefore the constitution of the lipid bilayer should be unique [8]. Fortunately, variant of SARS-CoV-2 has not been reported. There will be mutations in mRNA virus, but length of mRNA will be remaining same, because largely altered mRNA will lose infectious function and thus will not survive. Therefore, we may rely on size measurement to detect SARS-CoV-2 efficiently and reliably.

Figure 10 shows possible application to SARS-CoV-2 detection. The figure is the case using laser crossing method. The cavity resonator type will be also applicable.

We may operate this system as follows.

(1) Without fluorescent labeling, we run machine and detect particles. When a person voices, the vocal cords vibrate and chop the mucous into micron droplets, which contains viruses if the person is infected. Closing the shutter, and we wash the droplets into the solution.

(2) Using a series of porous filters, we remove micron-size particles, such as, pollen, dusts. At the same time, aggregated viruses will be isolated.

(3) If the histogram of the detected particle size does not show a peak around 100 nm (typical size of SARS-CoV-2), it is "NEGATIVE". Most of the cases are here.

(4) If the histogram has a peak, it is "NOT- NEGATIVE", we run the machine again.

We apply fluorescent labeling to detect spike protein of corona virus.

（5）If histogram measurement by fluorescent light shows "peak", it is "POSITIVE", if not it is "NEGATIVE".

If a PCR testing center is located on site, we may run this machine without fluorescent labeling and in case of "NOT-NEGATIVE", we may send this specimen to PCR testing. In this way, we may save washing time and cost for fluorescent labeling assay. By introducing parallel machine configuration, we may reduce the operation time less than 10 second, and cleaning time may be longer for each unit.

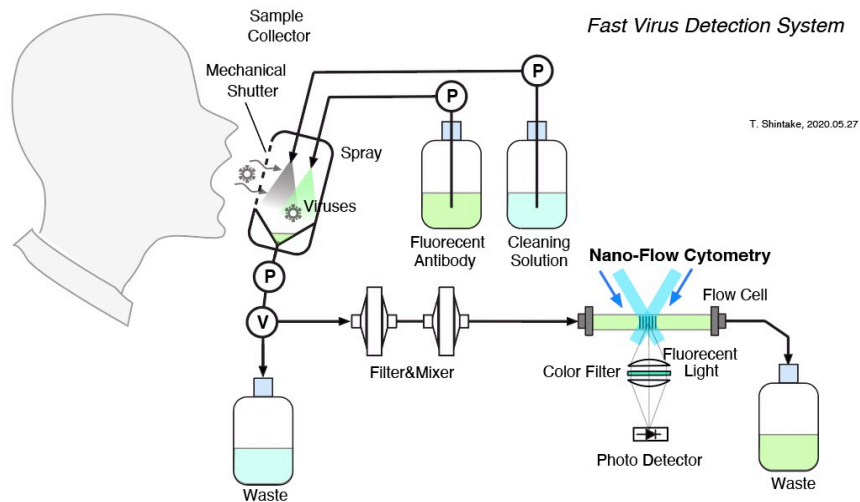

Fig. 12. Fast virus detection system based on flow cytometry.

## 8. Conclusions

In this paper, the author proposed the flow cytometry using laser interferometry. It will provide calibrated scale on the particle size measurement, more scattering photons from longer interaction length, and higher S/N due to high frequency modulation signal. The author hopes, the study on extracellular vesicles (EVs) and viruses of nanometer scale will be accelerated by the cytometry. And the fast virus detection system may make it possible to keep open the international airports, even under pandemic, by removing risk of spreading respiratory infectious diseases.

# References

[1] Marco Rieseberg, Cornelia Kasper, Kenneth F. Reardon, Thomas Scheper, "Flow cytometry in biotechnology", Appl Microbiol Biotechnol (2001) 56:350–360 DOI 10.1007/s002530100673

[2] Frank A.W. Coumans, "Methodological Guidelines to Study Extracellular Vesicles", Circulation Research 2017;120:1632-1648. DOI: 10.1161/CIRCRESAHA.117.309417

[3] Zoltán Vargaa et. al., "Size Measurement of Extracellular Vesicles and Synthetic Liposomes: The Impact of the Hydration Shell and the Protein Corona", Colloids and Surfaces B: Biointerfaces 192 (2020) 111053

[4] J. Lizbeth Reyes Zamora and Hector C. Aguilar, "Flow virometry as a tool to study viruses", Methods 134-135 (2018) 87-97

[5] Ye Tian et. al., "Protein Profiling and Sizing of Extracellular Vesicles from Colorectal Cancer Patients via Flow Cytometry", ACS Nano 2018, 12, 671−680, DOI: 10.1021/acsnano.7b07782

[6] E. VAN DER POL et. al., "Innovation in detection of microparticles and exosomes", Journal of Thrombosis and Haemostasis, 11 (Suppl. 1): 36–45, (2013), DOI: 10.1111/jth.12254

[7] W. G. Eisert, "High Resolution Optics Combined with High Spatial Reproducibility in Flow", CYTOMETRY Vol. 1, No. 4, pp. 254-259, 1981

[8] Paul S. Masters, "THE MOLECULAR BIOLOGY OF CORONAVIRUSES", ADVANCES IN VIRUS RESEARCH, VOL 66, 2006, DOI: 10.1016/S0065-3527(06)66005-3

[9] T. Shintake, "Proposal of a nanometer beam size monitor for e+e- linear colliders", Nuclear Instrument and Methods in Physics Research A311 (1992) 453-464

[10] Peter Tenenbaum and Tsumoru Shintake, "Measurement of Small Electron-Beam Spots", Annu. Rev. Nucl. Part. Sci. 1999. 49:125–62

[11] Jacqueline Yan, et. al., "Shintake Monitor, Nanometer Beam Size Measurement and Beam Tuning", Physics Procedia 37 (2012) 1989-1996.

## Appendix 1. Derivation of the modulation depth.

The scattering power from a particle inside the interference fringe is given by

$$P_s = \iint_S \sigma_s \, p_0(x,z) \, dx dz = \iint_S \sigma_s \frac{E^2}{\zeta} dx dz$$

$$= \iint_S \sigma_s \frac{E_0^2}{\zeta} cos^2(k_x x) \, dx dz$$

$$= \sigma_s \frac{P_L}{\sigma_L} \iint_S cos^2(k_x x) \, dx dz$$

$$= \sigma_s \frac{P_L}{\sigma_L} \iint_S \frac{1 + cos(2 k_x x)}{2} dx dz$$

$$= \sigma_s \frac{P_L}{\sigma_L} \int_{z_1}^{z_2} dz \int_{x_0 - d/2}^{x_0 + d/2} \frac{1 + cos(2k_x x)}{2} dx \quad \text{(A1-1)}$$

Where $P_L$ is laser power, $\sigma_L$ is laser spot size (cross-section), $x_0$ is the position of particle, $k_x$ is wave number in x-direction: $k_x p = \pi$, $z_1$ and $z_2$ is the integral limits representing circular particle. Equation a1 represents intensity modulation in the scattering light, which takes maximum at $2k_x x_0 = 0$, and minimum at $2k_x x_0 = \pm \pi$,

$$P_{max} = \sigma_s \frac{P_L}{\sigma_L} \int_{z_1}^{z_2} dz \int_{-d/2}^{+d/2} \frac{1 + cos(2k_x x)}{2} dx \quad \text{(A1-2)}$$

$$P_{min} = \sigma_s \frac{P_L}{\sigma_L} \int_{z_1}^{z_2} dz \int_{-d/2}^{+d/2} \frac{1 - cos(2k_x x)}{2} dx \quad \text{(A1-3)}$$

$$z_1 = -\sqrt{(d/2)^2 - x^2}$$

$$z_2 = +\sqrt{(d/2)^2 + x^2}$$

Using symmetry condition,

$$P_{max} = \sigma_s \frac{P_L}{\sigma_L} \int_0^{+d/2} 2\sqrt{(d/2)^2 - x^2} \, \{1 + cos(2k_x x)\} dx \quad \text{(A1-4)}$$

We approximately integrate this equation. The first term can be approximated as

$$2\sqrt{(d/2)^2 - x^2} \approx d \, cos(k_2 x) \quad \text{(A1-5)}$$

where $k_2 d = \pi$.

$$P_{max} = \sigma_s \frac{P_L}{\sigma_L} \int_0^{+d/2} d \, cos(k_2 x) \, \{1 + cos(2k_x x)\} dx$$

$$= \sigma_s d \frac{P_L}{\sigma_L} \int_0^{+d/2} \{cos(k_2 x) + cos(k_2 x) cos(2k_x x)\} dx$$

$$= \sigma_s d \frac{P_L}{\sigma_L} \left[ \frac{sin(k_2 x)}{k_2} + \frac{k_2}{k_2^2 - (2k_x)^2} sin(k_2 x) cos(2k_x x) + \frac{2k_x}{k_2^2 - (2k_x)^2} cos(k_2 x) sin(2k_x x) \right]_0^{d/2}$$

$$= \sigma_s d \frac{P_L}{\sigma_L} \left( \frac{1}{k_2} + \frac{k_2}{k_2^2 - (2k_x)^2} cos(k_x d) \right) \tag{A1-6}$$

We can find the same equation for minimum power with negative sign on the cosine term. Here we define modulation depth:

$$M = \frac{P_{max} - P_{min}}{P_{max} + P_{min}} \tag{A1-7}$$

$$M = \frac{k_2^2}{k_2^2 - (2k_x)^2} cos(k_x d)$$

$$= \frac{1}{1 - (2d/p)^2} cos(\pi d/p) \tag{A1-8}$$

### Appendix -2. Prior Art: "Shintake Monitor"

Previously, the laser interference method was successfully used to measure the electron beam size in high-energy particle accelerator, which is now called "Shintake Monitor". Firstly, the transverse size around 60 nm of 50 GeV electron beam was measured by using YAG laser of 1064 nm wavelength at FFTB experiment of SLAC Stanford in 1990s[9,10], and 30 nm spot size was measured by using second harmonics of YAG-laser: 532 nm wavelength at ATF facility KEK Japan in 2000s[11].

Figure A1 shows the schematic diagram of Shintake Monitor. The YAG-laser emits infra-red light at 1064 nm wavelength, followed by splitting by the half-mirror into two beams, and guided to the common interaction point inside the vacuum pipe of the high energy accelerator. Overlapping two laser beams creates periodic interference fringes, whose pitch: $p$ is given by

$$p = \frac{\lambda_0}{2 sin(\theta/2)} \tag{A2-1}$$

Where $\theta$ is the crossing angle, $\lambda_0$ is the wavelength of the laser. By using head on crossing, we obtain minimum pitch size of half-wavelength of laser. On the other hand, by using small angle, we may go much larger pitch size than the wavelength. Therefore, by choosing the crossing angle properly we may adjust sensitive range to the target size.

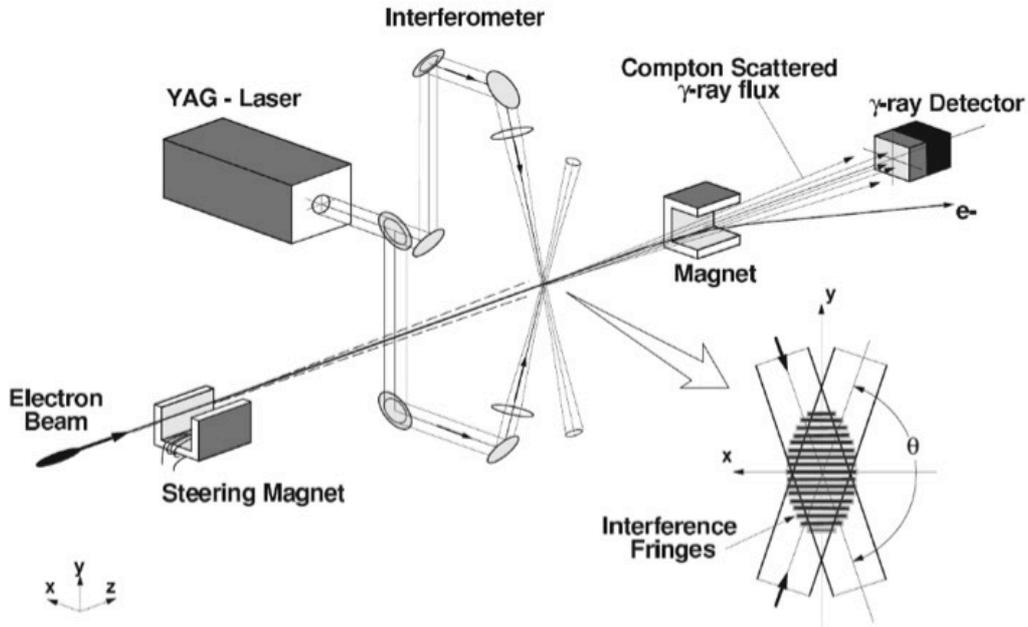

Fig. A1 System diagram of the spot size monitor.

When the high energy electron beam runs through the interaction point, some of the electron will collide to the photons of the laser and generate Compton scattering gamma-rays into the detector at downstream. We measured the gamma-ray intensity by scanning the vertical position of the electron beam at the interaction point.

As shown in Fig. A2, when the electron beam size is much smaller than the fringe pitch, the number of gamma-ray will vary as the light intensity in the fringe. If the electron beam size becomes larger, contributions form high and low zones cancel out, and the modulation depth becomes low. From the modulation measurement, we may obtain the spot size.

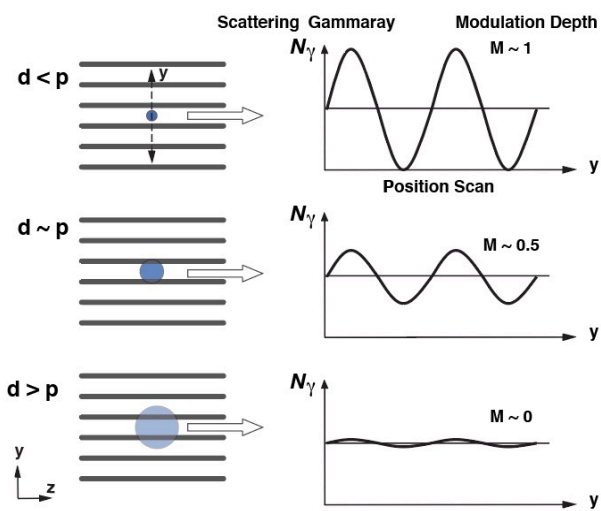

Fig. A2. Modulation depth depends on spot size.

The probability of colliding an electron and to a photon is given by the radiation coupling. We can treat this process by square of the electric field intensity of the laser field.

$$N_\gamma \sim E^2 = E_0^2 \, (1 + \cos\theta \cos 2k_y y)$$

We have to note that at the normal crossing $\theta = 90$ degree the modulation becomes zero, and thus this method does not work. This is due to the contributions from the electric field and magnetic field cancels each other's. In flow cytometry, speed of particle is much slower than the speed of light, thus magnetic field contribution becomes negligibly small, and it still works at normal crossing $\theta = 90$ degree.

Assuming Gaussian density distribution for the electron beam, we can estimate the number of gamma-ray as follows.

$$N(y_0) = \int_{-\infty}^{\infty} \frac{1}{\sqrt{2\pi}\sigma} e^{-(y-y_0)^2/2\sigma^2}(1 + \cos\theta \cos 2k_y y) \, dy$$

where $N(y_0)$ is arbitrarily normalized. $y_0$ is the vertical position of the electron beam. $k_y$ is the wave number in vertical direction. This is Fourier transform, i.e., $N(y_0)$ is the Fourier component of Gaussian density distribution at special frequency of $2k_y$.

$$2k_y p = 2\pi, \; p = \frac{\lambda_0}{2\sin(\theta/2)}$$

The modulation depth, defined by the peak-to-average divided by the average is

$$M = \frac{\delta N}{N} = |\cos\theta| \exp[-\frac{1}{2}(2k_y\sigma)^2]$$

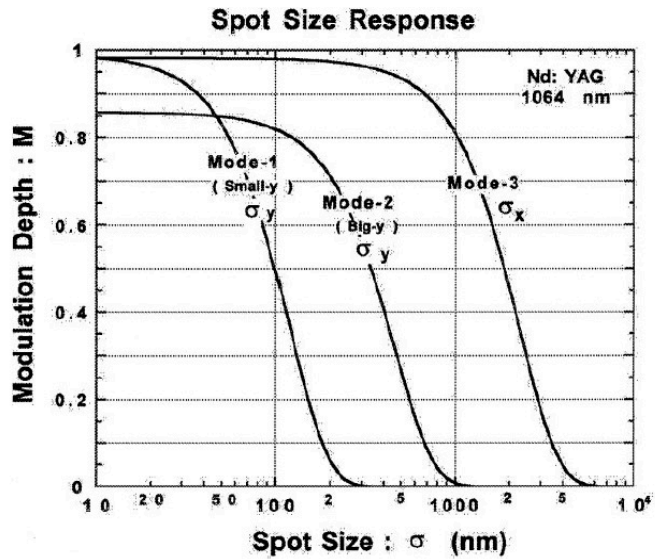

Fig. A3. Modulation depth as a function of beam size: the spot size monitor for Final Focus Test Beam: crossing angles of 174, 30 and 6 degree are shown.

Fig. A4 shows example of the measured data, which shows clear periodic variation associated with the laser interference fringe. From the modulation depth, we obtained the vertical spot size of the electron beam.

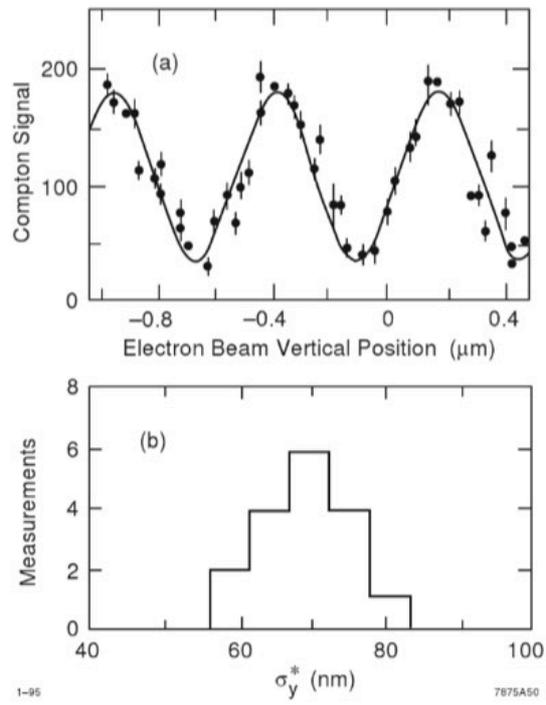

Fig. A4. Example measurement data at FFTB experiment.